\documentclass[aps,prl,twocolumn,groupedaddress,twocolumn,showpacs]{revtex4}

\usepackage{graphicx}
\usepackage{bm}

\begin{document}


\title{Measurement of local dissipation scales in turbulent pipe flow}


\author{S. C. C. Bailey$^1$, M. Hultmark$^1$, J. Schumacher$^2$,  V. Yakhot$^3$, and A. J. Smits$^1$}
\affiliation{$^1$Department of Mechanical and Aerospace Engineering, Princeton University, Princeton, NJ, 08540, USA,\\
$^2$ Department of Mechanical Engineering, Technische Universit\"{a}t Ilmenau, D-98684 Ilmenau, Germany,\\
$^3$ Department of Mechanical Engineering, Boston University, Boston, MA, 02215, USA}

\date{\today}

\begin{abstract}
Local dissipation scales are a manifestation of the intermittent small-scale nature of turbulence. We report the 
first experimental evaluation of the distribution of local dissipation scales in turbulent pipe flows for a range of 
Reynolds numbers, $2.4\times 10^4\leq Re_D \leq7.0\times10^4$.  Our measurements at the nearly isotropic
pipe centerline and within the anisotropic logarithmic layer show excellent agreement with distributions 
that were previously calculated from numerical simulations of homogeneous isotropic box turbulence and 
with those predicted by theory. The reported results suggest a universality of the smallest-scale fluctuations around
the classical Kolmogorov dissipation length.
\end{abstract}

\pacs{47.27.eb, 47.27.nf}

\maketitle


The dissipation of turbulent kinetic energy is driven by fluctuations of velocity gradients whose amplitudes 
can locally and instantaneously greatly exceed their mean value, resulting in spatially intermittent regions of high rates 
of turbulent dissipation within a turbulent flow field \cite{Sreenivasan1997,Kaneda2009}. The statistical mean 
of the turbulent kinetic energy dissipation rate equals the average flux of energy
from the energy-containing large-scale eddies down to the smallest ones in case of statistically stationary 
turbulent fluid motion \cite{Kolmogorov1941}. In the classical theory of turbulence, this small-scale end of 
the turbulent cascade is then prescribed by the Kolmogorov dissipation scale $\eta_K$ which is derived as $\eta_K=\nu^{3/4}/\left\langle \epsilon \right\rangle ^{1/4}$,
where $\nu$ is the kinematic viscosity of the fluid and $\langle \epsilon \rangle$ is the mean of the energy
dissipation rate field which is given by $\epsilon({\bm x},t)=(\nu/2)(\partial_i u_j+\partial_j u_i)^2$. But
is $\eta_K$ the smallest scale in a turbulent flow? The Kolmogorov dissipation length contains 
the mean dissipation rate $\langle\epsilon\rangle$ and does not consider the strongly intermittent nature of 
$\epsilon({\bm x},t)$. It seems therefore natural to capture for these fluctuations in a dissipation scale 
definition and to refine the notion of one {\em mean} dissipation length to that of a whole continuum of 
{\em local} dissipation scales. The finest of those scales will then be associated with the steepest 
velocity gradients, or alternatively, with the highest-amplitude shear and
vorticity events in turbulence.  This was done first within the multifractal formalism and demonstrated that
local scales below $\eta_K$ will exist \cite{Paladin1987,Nelkin1990,Frisch1991,Biferale2008}. An alternative 
approach to such a continuum of dissipation scales was suggested by Yakhot \cite{Yakhot2005,Yakhot2006}. 
When connecting a local scale $\eta$ and the velocity increment across that scale,
$\delta_{\eta} u = | u(x+\eta)-u(x) |$, the following relation 
\begin{equation}
\label{eq_3}
\eta\delta_{\eta} u \approx \nu
\end{equation}
was obtained from the Navier-Stokes equations of fluid motion by a so-called point-splitting technique. 
Put into another perspective, a {\em local}  Reynolds number obeying this scale and velocity as their 
characterisitic amplitudes will be $Re_{\eta} = \eta\delta_{\eta} u/ \nu = O(1)$. Such a Reynolds 
number is then also associated with fluid motion at the crossover scales between the inertial cascade 
range and the viscous dissipation range. 

The probability density function (PDF) of the local dissipation 
scales, $Q(\eta)$, contains scales larger and smaller than the classical Kolmogorov dissipation length 
and can be calculated  analytically on the basis of these ideas \cite{Yakhot2006}. The analytic result
was found recently to be in good agreement with a direct calculation from high-resolution numerical 
simulation data of three-dimensional homogeneous isotropic box turbulence 
\cite{Schumacher2007,Schumacher2007a}.  The numerical simulations in \cite{Schumacher2007} were 
conducted for relatively low Taylor-microscale Reynolds numbers ($14\le Re_{\lambda}\le 151$, where $Re_\lambda=u_x'\lambda/\nu$, $\lambda=\sqrt{2(15\nu u_x'^2)/\langle\epsilon\rangle}$ and the standard deviation of the turbulent velocity component is $u'_x$) lacking 
even traces of inertial range. Still the scaling of moments of velocity gradients with Reynolds number agreed
with the predictions which are obtained when using the scaling exponents of inertial range structure functions 
for high-Reynolds number flows. This somewhat unexpected result suggested that the existence of a 
well-developed inertial range might not be essential for achieving an asymptotic regime of the small-scale gradient 
statistics. 

In this Letter, we want to advance these ideas into two directions. We report the first experimental 
confirmation of the local dissipation scale concept in a laboratory flow. Furthermore, we demonstrate 
the robustness of the concept beyond the idealized situation of homogeneous isotropic box turbulence. The
flow at hand is a turbulent pipe flow in which a mean shear and thus an inhomogeneous (radial) direction
are present.  Interestingly, we can also report that the monitored dissipation scale distributions at the nearly isotropic pipe centerline and in the strongly anisotropic logarithmic layer are almost identical for the range of  
Reynolds numbers that is spanned by the experiment. Additionally, they are in very good agreement 
with those of the numerical simulations \cite{Schumacher2007a} and the analytical result \cite{Yakhot2006}. 
Our experiments span an Reynolds number range of $76\leq Re_{\lambda} \leq 214$ .  

{\em Description of the pipe flow experiment.}
The experiments were conducted using the Princeton University/ONR Superpipe \cite{Zagarola1998}, which consists of a closed return-pressure vessel containing a long test pipe downstream of flow-conditioning and heat-exchanging sections. The current set of measurements were performed using air at room temperature and atmospheric pressure. The test section used during these experiments consisted of commercial steel pipe \cite{Langelandsvik2008} with an average inner radius of $R=64.9\textrm{mm}$ and an overall length of $200D$. Experiments were performed over an interval of pipe flow Reynolds numbers (based on pipe diameter $D$ and area averaged velocity $U_B$) of 
$2.4\times10^4 \leq Re_D=U_BD/\nu \leq 7.0\times10^4$. As detailed in \cite{Langelandsvik2008,Zagarola1998}, over this Reynolds number range, the flow at the test section of the pipe was fully developed, smooth walled turbulent pipe flow.  

Velocity measurements of streamwise velocity, $U_x$, were performed with a single sensor hot-wire probe $2.5\mu\textrm{m}$ in diameter and having a sensing length of $l=0.5\textrm{mm}$.  The wire was operated at an overheat ratio of 1.8. Analog output from the anemometer was low-pass filtered at half the sample frequency using an external filter before being digitized using a 16-bit simultaneous sample and hold A/D board.   Sampling rate was altered between 100kHz and 200kHz in order to maximize the temporal resolution while not exceeding the measured frequency response of the probe.  Data was sampled for 30 minutes continuously. During this period, the flow temperature was found to remain relatively constant, changing by less than $1^{\circ}C$.

Hot-wire probe calibrations were performed at the pipe centerline using a Pitot-probe/wall pressure tap combination.  Calibrations were performed before and after each 30 minute measurement.  Data sets where voltage drift was observed were discarded.

Measurements were performed at two positions, the first located at the pipe centerline, and the second within the logarithmic layer at a distance from the wall of $y = 0.1R$.  The flow conditions for each measurement are provided in Tables \ref{CL_conditions} and \ref{yR0p1_conditions} for the $y/R = 1$ and $y/R = 0.1$ measurements respectively.

 \begin{table}
 \caption{\label{CL_conditions} Experiment conditions for measurements performed at $y/R=1$. $\overline{U}_x$ is the time averaged streamwise velocity.}
 \begin{ruledtabular}
 \begin{tabular}{cccccccc}
$Re_D$ & $Re_L$ & $Re_\lambda$ & $\overline{U}_x \textrm{(m/s)}$ & $u_x' \textrm{(m/s)}$ &$L \textrm{(m)}$ & $\eta_K \textrm{(mm)}$ & $\langle\epsilon\rangle (\textrm{m}^2/\textrm{s}^2)$\\
\hline
24000 & 230 & 76 & 3.3 & 0.12 &0.029  &0.47 & 0.07 \\
28000 & 290 & 87 & 3.9 & 0.15 &0.029  &0.40 & 0.13 \\
35000 & 340 & 92 & 4.9 & 0.17 &0.029  &0.35 & 0.21 \\
44000 & 440 & 106 & 6.0 & 0.21 &0.030  &0.31 & 0.39 \\
52000 & 560 & 116 & 7.2 & 0.25 &0.033  &0.27 & 0.62 \\
60000 & 700 & 124 & 8.3 & 0.29 &0.036  &0.25 & 0.90 \\
70000 & 780 & 135 & 9.5 & 0.33 &0.036  &0.22 & 1.31 \\
 \end{tabular}
 \end{ruledtabular}
 \end{table}
\begin{table}
 \caption{\label{yR0p1_conditions} Experiment conditions for measurements performed at $y/R=0.1$.}
 \begin{ruledtabular}
 \begin{tabular}{cccccccc}
$Re_D$ & $Re_L$ & $Re_\lambda$ & $\overline{U}_x \textrm{(m/s)}$ & $u_x' \textrm{(m/s)}$ &$L \textrm{(m)}$ & $\eta_K \textrm{(mm)}$ & $\langle\epsilon\rangle (\textrm{m}^2/\textrm{s}^2)$\\
\hline
35000 & 1300 & 155 & 3.4 & 0.41 & 0.047 & 0.19  & 2.4 \\
52000 & 1900 & 185 & 5.1 & 0.59 & 0.050 & 0.15  & 6.8 \\
70000 & 2700 & 214 & 6.7 & 0.76 & 0.055 & 0.13  & 14  \\
\end{tabular}
 \end{ruledtabular}
 \end{table}
\begin{table}
 \caption{\label{confidence} Minimum values of $\eta$ which can be considered fully free 
of effects of $\eta_l$ sensor spatial resolution and $\eta_f$ signal noise (defined using the 
wavenumber where the signal to noise ratio exceeded 10). Scales are also shown normalized by $\eta_0=L Re_L^{0.72}$.}
 \begin{ruledtabular}
 \begin{tabular}{ccccc}
$Re_\lambda$ & $\eta_l \textrm{(mm)}$ & $\eta_l / \eta_0$ & $\eta_f \textrm{(mm)}$ & $\eta_f / \eta_0$\\
\hline
76 & 0.5 & 1.0 & 1.4  & 2.9 \\
87 & 0.5 & 1.2 & 1.4  & 3.5 \\
92 & 0.5 & 1.4 & 1.5  & 4.1 \\
106 & 0.5 & 1.6 & 1.3  & 4.2 \\
116 & 0.5 & 1.7 & 1.3  & 4.7 \\
124 & 0.5 & 1.9 & 1.4  & 5.2 \\
135 & 0.5 & 2.0 & 1.4  & 5.8 \\
155 & 0.5 & 2.2 & 1.0 & 4.5 \\
185 & 0.5 & 2.8 & 0.7 & 3.8 \\
214 & 0.5 & 3.2 & 0.5 & 3.0 \\
 \end{tabular}
 \end{ruledtabular}
 \end{table}

{\em Confidence limits of the measurements.}
When resolving the small scale structures of turbulence using hot-wire anemometry, the limitation imposed by the finite sensor length of the probe is an important consideration.  Insufficient spatial resolution will filter the energy at scales smaller than the probe.  For the current measurements, the probe sensor length was of the order of the Kolmogorov scale with a maximum of  $l\approx 4\eta_K$.  Although it has been observed that the energy spectrum can be well resolved for turbulence scales up to $0.05l$ \cite{Ligrani1987}, the impact of spatial filtering on the instantaneous velocity is not well understood, thus full confidence can only be given for measurements of $\eta$ for scales larger than $l$.

\begin{figure}
\includegraphics[width=3.375in]{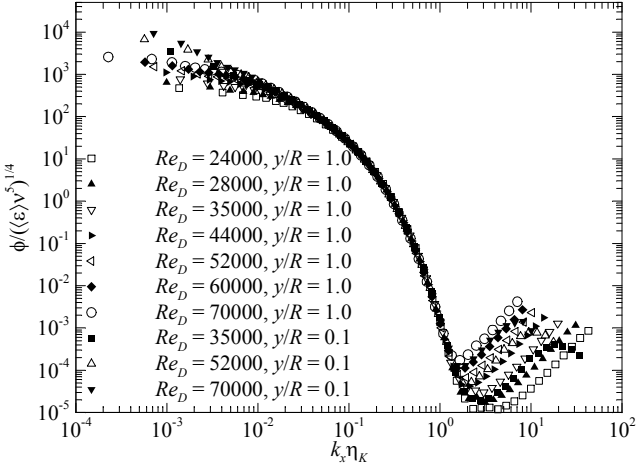}
\caption{\label{spectra}One-dimensional turbulence energy spectra measured for all cases. The legend
indicates whether the data are obtained at the centerline $y/R=1.0$ or in the logarithmic layer $y/R=0.1$.}
\label{fig1}
\end{figure}

The scaled one-dimensional energy spectra, $\phi(k_x)$, are plotted in Fig. \ref{spectra} for the centerline, 
$y/R=1.0$, and in the logarithmic layer, $y/R = 0.1$. Streamwise wavenumber, $k_x$, was estimated from 
frequency, $f$, using Taylor's frozen flow hypothesis through $2\pi f / \overline{U}_x$.  As can be observed in 
Fig. \ref{spectra}, the signal to noise ratio fell below one (as indicated by the inflection point in the measured spectra) at wavenumbers larger than the Kolmogorov range for all data sets.  Note that the appearance of this noise is consistent with the so called $f^2$ noise typically observed in constant temperature anemometry.  

\begin{figure}
\includegraphics[width=3.375in]{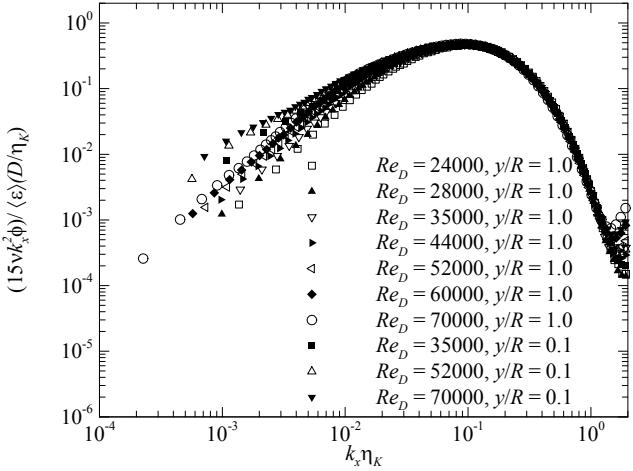}
\caption{\label{diss_spectra}Measured dissipation spectrum for all cases. Data are the same as in Fig. 
\ref{fig1}.}
\end{figure}
 
\begin{figure}
\includegraphics[width=3.375in]{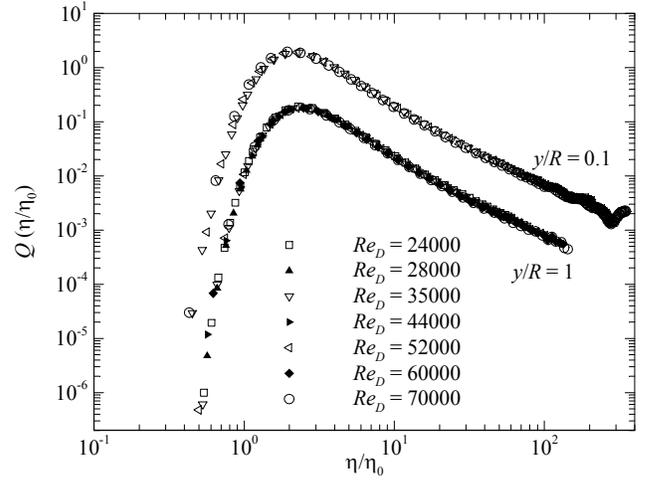}
\caption{\label{diss_pdfs} Measured PDFs of the local dissipation scales for all Reynolds numbers.  
Results from the logarithmic layer $y/R=0.1$ are shifted vertically by a decade for clarity.}
\end{figure}

An isotropic estimate of the mean dissipation rate can be found through integration of the one-dimensional dissipation spectrum
\begin{equation}
\langle\epsilon\rangle=15 \nu \int_0^\infty \!  k_x^2 \phi(k_x) \, dk_x.
\end{equation}
The measured dissipation spectra, shown in Fig. \ref{diss_spectra}, indicate that the range of wavenumbers over which dissipation occurred were well resolved during these measurements.  Thus, the measured PDFs are expected to be largely insensitive to the influence of spatial filtering and instrumentation noise. The insensitivity of the results to changing Reynolds number and $\eta_K$ in Fig. \ref{diss_pdfs} confirm this expectation. Full confidence, however, can only be given to results for values of $\eta$ larger than the scales at which these effects are expected to occur.  These minimum confidence limits are quantified in Table \ref{confidence}.

{\em Distribution of local dissipation scales.}
The probability density function (PDF) of $\eta$, $Q(\eta)$ can then be calculated by a saddle-point approximation
\cite{Yakhot2006} within the range  $0<\eta<L$ as 
\begin{eqnarray}
\label{eq_5}
Q(\eta) & = & \frac{1}{\pi\eta\left(b \log\left(L/\eta\right)\right)^{1/2}}\int_{-\infty}^\infty{dx} \nonumber \\
& & \times \exp\left[ -x^2- \frac{\left(\log\left(\sqrt{2}xRe_L\left(\eta/L\right)^{a+1}\right)\right)^2}{4b\log\left(L/\eta\right)}\right]
\end{eqnarray}
where $a=0.383$, $b=0.0166 $ \cite{Schumacher2007a}, $L$ is the integral length scale of the turbulence and $Re_L=\langle|u_x(x+L)-u_x(x)|^2\rangle^{1/2} L / \nu$ as used in \cite{Schumacher2007a}.
The PDF was calculated from each velocity time series using the following procedure.  First, the velocity difference  $\Delta u(\Delta t) = |U_x(t+ \Delta t)-U_x(t)|$ was calculated at each discrete time, $t$.  Next, the value $\Delta u(\Delta t)\overline{U}_x \Delta t/\nu$ was calculated and the instances where the result was between 0.9 and 2 were counted as occurrences of dissipation at a scale $\eta = \Delta t \overline{U}_x$.  This process was repeated for all $\Delta t$ up to  $\Delta t \overline{U}_x=L$.  Finally	$Q(\eta)$ was found from the count of occurrences by normalizing such that $\int Q(\eta)d\eta=1$.

\begin{figure}
\includegraphics[width=3.375in]{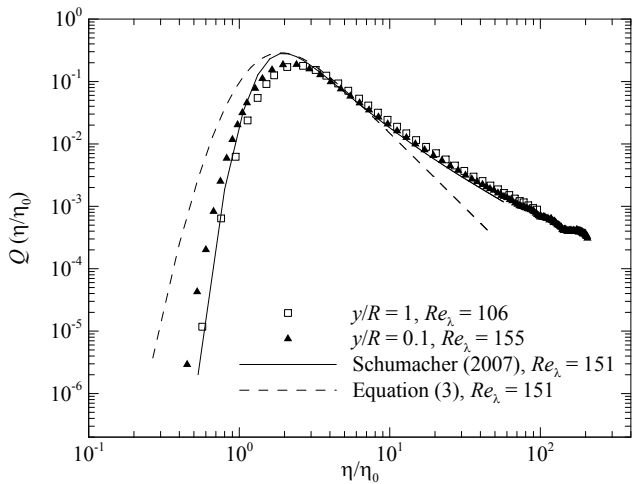}
\caption{\label{diss_comp} Comparison between $Q(\eta/\eta_0)$ measured at pipe centerline, $y/R=1.0$, measured at $y/R=0.1$, numerical results of \cite{Schumacher2007a} and theoretical distribution of \cite{Yakhot2006}.}
\end{figure}

The results are shown in Fig. \ref{diss_pdfs}, normalized by the value $\eta_0=LRe_L^{-0.72}$ \cite{Schumacher2007a}, which is very close to $\eta_K$.  At each measurement location there was excellent collapse of $Q(\eta / \eta_0)$ for all measured Reynolds numbers.  This Reynolds number independence indicates that the high-wavenumber end of the universal equilibrium range could establish universality at lower Reynolds numbers than previously expected.  This appears to be true even though the scale separation between the energy containing eddies and dissipation scales is still relatively small and before even the establishment of an inertial subrange (see Fig. \ref{spectra}).

Direct comparison is made in Fig. \ref{diss_comp} between $Q(\eta/\eta_0)$ measured in the 
nearly isotropic turbulence at the pipe centerline and the anisotropic log-layer. Here, one can 
see nearly identical distributions at all but the smallest values of $\eta/\eta_0$ where a slightly 
increased probability was observed at $y/R=0.1$.  Such a good agreement indicates that the 
structure of turbulence at the largest scales has little effect on the organization of the smallest 
scales.  

Comparison is also made in Fig. \ref{diss_comp} between the present experimental 
results, the numerical results of \cite{Schumacher2007a} and the theoretical distribution of \cite{Yakhot2006}.  
There is a very good agreement between the numerical and experimental results (at the center line), 
also in the tails of the PDF at large $\eta$. The deviations from the theoretical distribution on both 
sides can be due to the limits of a saddle-point approximation for the evaluation of the Mellin 
transform in \cite{Yakhot2006}. The agreement between the numerical and experimental results 
supports the concept of the universality of the distribution of the dissipation scales.   

As mentioned above, Fig. \ref{diss_comp} displays a further property. The left tail of the PDF $Q(\eta/\eta_0)$ becomes 
slightly fatter for the logarithmic layer in comparison to that of the center line. Since we found that the
data for different Reynolds numbers collapse well (see Fig. \ref{diss_pdfs}), this property could be 
interpreted as a statistical fingerprint of bursting structures.  So-called packets of hairpins have been 
observed recently in the logarithmic layer of other wall flows \cite{Adrian2007}.  They were found to 
increase the level of small-scale intermittency which would be in accordance with the slightly smaller measured local dissipation scales. 
However, a clear disentanglement requires two things, further data records at higher $Re$ combined with experiments 
in different flows.

{\em Conclusions.}
The probability density function of the local dissipation scales was measured in a turbulent pipe flow 
over a range of Reynolds numbers at the pipe centerline and in the logarithmic layer. For the first time, 
we measured this probability density function experimentally in a turbulent shear flow.  Our results indicate 
that the distribution is basically independent of both the Reynolds number and the degree of anisotropy of the 
large scales of turbulence. We find a very good agreement with  the numerical simulations and the theory.
This robustness of the results with respect to flow-type (or degree of anisotropy)  suggests a universality of 
the smallest-scale fluctuations in turbulence. One could thus indeed conclude that the turbulent 
dynamics  at the finest scales is already in an asymptotic state although the Reynolds numbers remain moderate
and an inertial cascade range is absent or very small.  It is well-known that to observe inertial range statistics 
in wall-bounded flows requires large Reynolds numbers \cite{Saddoughi1994}. This fact
adds further value to the present study which establishes a connection between shear flow and isotropic 
turbulence at their small scales.

\begin{acknowledgments}
We wish to thank K. R. Sreenivasan for helpful suggestions and comments. 
S.B., M.H. and A.S. are supported by the Office of Naval Research (ONR) under Grant N00014-07-1-0111. Support for S.B. was provided by the PDF program of the Natural Sciences and Engineering Research Council (NSERC). J.S. is supported by the Heisenberg Program of the Deutsche Forschungsgemeinschaft (DFG) under Grant SCHU1410/5-1. 
\end{acknowledgments}

\end{document}